%% file: main.tex
\documentclass{article}

\usepackage{microtype}
\usepackage{graphicx}
\usepackage{subcaption}
\usepackage{booktabs} % for professional tables

\usepackage{hyperref}

\usepackage[preprint]{icml2026}

\usepackage{amsmath}
\usepackage{amssymb}
\usepackage{mathtools}
\usepackage{amsthm}

\usepackage[capitalize,noabbrev]{cleveref}

\theoremstyle{plain}

\theoremstyle{definition}

\theoremstyle{remark}

\usepackage[textsize=tiny]{todonotes}

\icmltitlerunning{Preserving Vertical Structure in 3D-to-2D Projection for Permafrost Thaw Mapping}

\begin{document}

\twocolumn[
  \icmltitle{Preserving Vertical Structure in 3D-to-2D Projection for Permafrost Thaw Mapping}



  \icmlsetsymbol{equal}{*}

  \begin{icmlauthorlist}
    \icmlauthor{Justin McMillen}{yyy}
    \icmlauthor{Robert Van Alphen}{xxx}
    \icmlauthor{Taha Sadeghi Chorsi}{xxx}
    \icmlauthor{Jason Shabaga}{ccc}

    \icmlauthor{Mel Rodgers}{xxx}
    \icmlauthor{Rocco Malservisi}{xxx}
    \icmlauthor{Timothy Dixon}{xxx}
    \icmlauthor{Yasin Yilmaz}{yyy}
    %\icmlauthor{Firstname8 Lastname8}{yyy,comp}
  \end{icmlauthorlist}

  \icmlaffiliation{yyy}{Electrical Engineering Department, University of South Florida, Tampa, FL, USA}
  \icmlaffiliation{xxx}{Geoscience Department, University of South Florida, Tampa, FL, USA}
  \icmlaffiliation{ccc}{XXX Department, University of Colorado Boulder, Boulder, CO, USA}

  \icmlcorrespondingauthor{Yasin Yilmaz}{yasiny@usf.edu}
  \icmlcorrespondingauthor{Timothy Dixon}{thd@usf.edu}

  \icmlkeywords{Machine Learning, ICML, Point Clouds, Regression, Classification}

  \vskip 0.3in
]



\printAffiliationsAndNotice{}  % no special notice (required even if empty)

\begin{abstract}
  Forecasting permafrost thaw from aerial lidar requires projecting 3D point cloud features onto 2D prediction grids, yet naive aggregation methods destroy the vertical structure critical in forest environments where ground, understory, and canopy carry distinct information about subsurface conditions. We propose a projection decoder with learned height embeddings that enable height-dependent feature transformations, allowing the network to differentiate ground-level signals from canopy returns. Combined with stratified sampling that ensures all forest strata remain represented, our approach preserves the vertical information critical for predicting subsurface conditions. Our approach pairs this decoder with a Point Transformer V3 encoder to predict dense thaw depth maps from drone-collected lidar over boreal forest in interior Alaska. Experiments demonstrate that z-stratified projection outperforms standard averaging-based methods, particularly in areas with complex vertical vegetation structure. Our method enables scalable, high-resolution monitoring of permafrost degradation from readily deployable UAV platforms.
\end{abstract}

\input{Sections/1_introduction}
\input{Sections/2_related_works}
\input{Sections/3_data_collection}
\input{Sections/4_machine_learning_methods}

\input{Sections/5_experiments}
\input{Sections/6_discussion}
\input{Sections/7_conclusion}

\bibliography{bibliography}
\bibliographystyle{icml2026}
\input{Sections/8_appendix}

%%%%%%%%%%%%%%%%%%%%%%%%%%%%%%%%%%%%%%%%%%%%%%%%%%%%%%%%%%%%%%%%%%%%%%%%%%%%%%%
%%%%%%%%%%%%%%%%%%%%%%%%%%%%%%%%%%%%%%%%%%%%%%%%%%%%%%%%%%%%%%%%%%%%%%%%%%%%%%%
% APPENDIX
%%%%%%%%%%%%%%%%%%%%%%%%%%%%%%%%%%%%%%%%%%%%%%%%%%%%%%%%%%%%%%%%%%%%%%%%%%%%%%%
%%%%%%%%%%%%%%%%%%%%%%%%%%%%%%%%%%%%%%%%%%%%%%%%%%%%%%%%%%%%%%%%%%%%%%%%%%%%%%%

%%%%%%%%%%%%%%%%%%%%%%%%%%%%%%%%%%%%%%%%%%%%%%%%%%%%%%%%%%%%%%%%%%%%%%%%%%%%%%%
%%%%%%%%%%%%%%%%%%%%%%%%%%%%%%%%%%%%%%%%%%%%%%%%%%%%%%%%%%%%%%%%%%%%%%%%%%%%%%%

\end{document}

%% file: Sections/1_introduction.tex
\section{Introduction}
Regions around the globe are facing environmental impacts from climate warming, including ecosystem shifts and infrastructure hazards. The circumpolar north has experienced widespread permafrost thaw throughout the last 50 years. Permafrost, defined as soil, rock, or organic material that remains frozen for at least two consecutive years \cite{Lewkowicz2025}, stores substantial organic carbon. This thawing can create a feedback loop, releasing more greenhouse gasses that intensify warming. 

The boreal forest of Alaska’s interior is one such place experiencing warming \cite{IPCC2021}. This region's discontinuous permafrost is controlled by interactions between climate, ecology, and hydrology. Boreal forests provide thick organic soils and canopies which allow permafrost to remain frozen at mean annual air temperatures above 0°C \cite{Bonan1989,Jorgenson2010,Zhu2019}. Disturbances such as warming temperatures or wildfires increase the likelihood of thaw initiation \cite{Camill1999,Yoshikawa2002,Jorgenson2022}, which can induce ground subsidence (thermokarst) that develops into sinkhole like features. This change in topography can expand laterally, changing boreal forest into wetlands, releasing green houses gases, and destabilizing infrastructure. \cite{Osterkamp2000,Haughton2018,vander2018,Dearborn2021}. 

There is a growing need for modeling which can detect and predict thermokarst induced surface deformation \cite{permafrost_collapse}. While traditional field measurements remain accurate, they are spatially limited \cite{permafrost_from_space, gtnp2015gtnf}. Airborne lidar offers high-resolution 3D coverage of terrain and vegetation \cite{Reutebuch01102003}, but the input is an unordered 3D point cloud, yet the desired output is a georeferenced 2D map suitable for analysis and decision-making.

Existing 3D-to-2D projection methods aggregate point features within fixed spatial bins, an approach suited to autonomous driving but problematic in forests \cite{li2023, hu2022}. Vertical structure carries predictive signal about subsurface conditions, yet standard aggregation oversamples dense canopy while underrepresenting sparse but informative ground returns \cite{vegetation_influence, Kropp_2021, rs2030833, CAMPBELL2018330}.

We propose a projection decoder with learned height embeddings that preserve vertical structure during 3D-to-2D transformation. Our method applies farthest point sampling in the z-dimension to ensure all forest strata are represented regardless of point density, then augments selected points with learned z-embeddings that enable height-dependent feature trhansformations. Combined with multi-scale feature fusion, our approach produces high-resolution thaw depth maps from single-date lidar.

We summarize our contributions as follows:
\begin{itemize}
    \item A learned height embedding that enables the network to apply height-dependent feature transformations during 3D-to-2D projection, allowing differentiation between ground-level and canopy-derived information.
    \item A multi-scale late fusion decoder that independently projects features from each encoder stage, capturing both fine detail and global context.
    \item Experimental analysis showing that explicit height encoding is the critical component for accurate thaw prediction, outperforming both naive aggregation and histogram baselines in both regression and classification formulations.
\end{itemize}

%% file: Sections/2_related_works.tex
\section{Related Work}

\subsection{Permafrost Monitoring and Forecasting}
The CALM project (Circumpolar Active Layer Monitoring) has conducted in-situ monitoring of permafrost for more than three decades \cite{Brown2000}. The project employs temperature monitoring, mechanical probing, and vertical displacement measurements to estimate active layer thickness (ALT), the layer that undergoes seasonal freeze-thaw.

Regional or global monitoring efforts expand on these ground-based measures with remotely sensed data \cite{Jorgenson, Kokelj}. Lidar, radar, electromagnetic, and spectral imagery have all added to the tools with which permafrost and its related environmental expressions can be studied \cite{Pastick2013,Li2015,SadeghiChorsi2024,Lu2025}. Unoccupied aerial vehicles (UAVs) have also established a role in permafrost and active layer studies due to their capacity to produce high-resolution digital surface models through photogrammetry \cite{Gaffey}. UAV-lidar, like photogrammetry, can produce point clouds and digital surface models, but where they differ is the ability for ground penetration \cite{Gao2024,Renette2024}. UAV-lidar can produce full vertical profiles from the ground surface up into the canopy producing a product with higher information density. In order to fully utilize such large datasets, machine learning models have become standard tools. These models have been used to extrapolate permafrost extent, detect thaw induced landslides, and predict active layer thickness \cite{Pastick2013,Li2017,Lou2023}.  

\subsection{3D Computer Vision}
Point cloud deep learning has evolved from per-point processing to transformer architectures capable of handling large-scale outdoor lidar. PointNet introduced learning directly on unstructured point sets using shared MLPs and symmetric pooling for permutation invariance \cite{PointNet}. PointNet++ extended this through hierarchical set abstraction layers with multi-scale grouping to handle non-uniform densities \cite{PointNet++}. The Point Transformer series introduced attention mechanisms for capturing long-range dependencies. Point Transformer applied vector attention within local neighborhoods \cite{PointTransformer}. Point Transformer V2 unified multi-head and vector attention with partition-based pooling \cite{PointTransformerV2}, and Point Transformer V3 prioritized scalability by replacing $k$-NN searches with serialized neighbor mapping and Flash Attention \cite{PointTransformerV3}. These efficiency gains make PTv3 suitable as an encoder backbone for processing large drone-collected point clouds.

\subsection{3D-to-2D Feature Projection}
The autonomous driving domain has developed bird's-eye-view (BEV) projection methods for dense prediction \cite{li2023}. VoxelNet introduced voxel feature encoding to transform points within spatial bins \cite{VoxelNet}, while PointPillars improved efficiency by organizing points into vertical pillars and using PointNets to learn per-pillar features scattered to pseudo-images \cite{PointPillars}. Camera-based approaches such as Lift-Splat-Shoot \cite{LiftSplatShoot} and BEVFormer \cite{BEVFormer} lift image features to 3D before aggregating to BEV grids. BEVFusion \cite{BEVFusion} unified lidar and camera modalities in a shared BEV space.

These BEV methods were designed primarily for object detection rather than continuous dense regression, and typically employ fixed spatial binning that allows dense regions to dominate aggregated features. Our approach instead queries $k$-nearest neighbors in XY for each output pixel, then applies farthest point sampling in Z to ensure vertical coverage. We then augment selected points with learned height embeddings, with learned height embeddings enabling height-dependent feature transformations.

%% file: Sections/3_data_collection.tex
\section{Data Collection}
\label{sec:data_collection}

\begin{figure}[t]
    \centering
    \includegraphics[width=\linewidth]{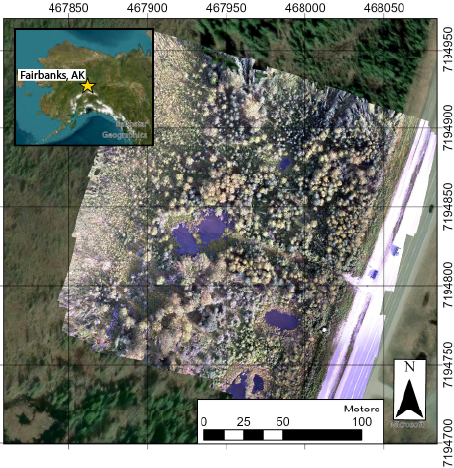}
    \caption{Orthophoto field map of Farmer's Loop 2 field site. Extent of the orhtophoto matches the extent of the lidar data. Inset: Location of Fairbanks, AK where our field site is located.}
    \label{fig:dataset}
    \vspace{-5mm}
\end{figure}

We chose a small field site operated by the U.S. Army Corps of Engineers' Cold Regions Research and Engineering Lab. This site (Farmer's Loop 2 (FL2)), located in Fairbanks, Alaska, is one of two thermokarst study areas on Farmer's Loop Road. This site was chosen as it contains several thermokarst indicators. Figure \ref{fig:dataset} shows both the site's location in Alaska, as well as a bird's eye view of the data collection site. It displays thermokarst related microtopographic shifts, changing vegetation cover, and hydrologic conditions indicative of rapidly thawing permafrost. Specifically, vegetation here is moving to a predominantly woody vegetation cover which can be indicative of warming soils along side recent measurements showing a deepening active layer.

\begin{figure}[t]
    \centering
    \includegraphics[width=\linewidth]{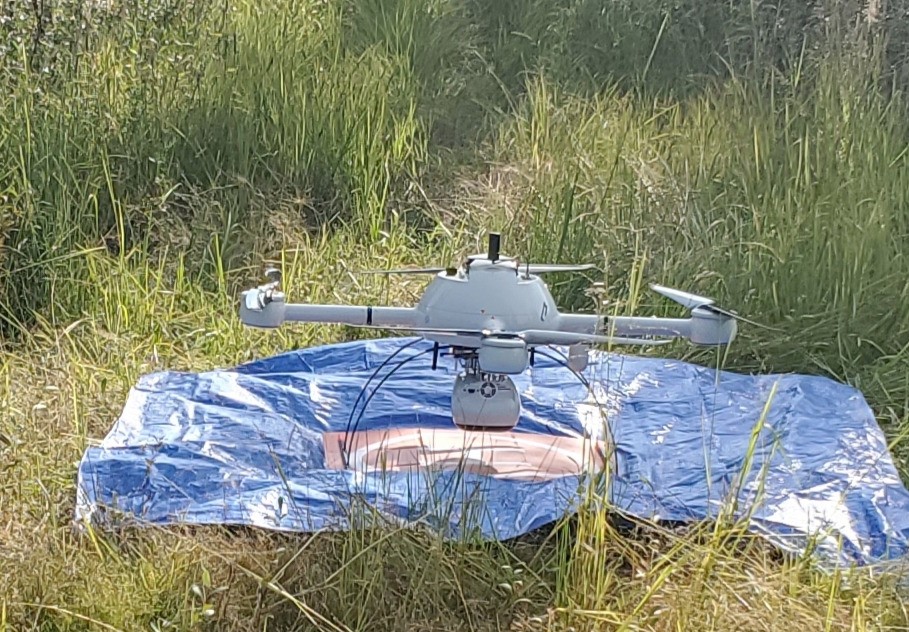}
    \caption{md-Lidar1000HR UAV with lidar and RGB camera.}
    \label{fig:drone}
    \vspace{-5mm}
\end{figure}

We collected data at FL2 at the end of freeze in May 2024 and at the end of summer thaw in August 2024 to capture the ground-truth thaw map. An md-Lidar1000HR UAV equipped with a small-form-factor dual lidar-camera system was used. The UAV is pictured in Figure \ref{fig:drone}. The UAV-lidar is a combination of a Velodyne Puck VLP-16 lidar and a SONY IMX264 camera. Images are captured simultaneously with the lidar data. The VLP-16 lidar outputs a near-infrared laser pulse (903 nm wavelength) at a repetition rate of approximately 290,000 pulses per second, with a scan angle of 15° on either side of nadir and a 360° horizontal field of view.  

To enable ground-truth comparison to the lidar point cloud we deployed check points throughout the field site. Checkpoint locations were selected to optimize for visibility to the UAV-lidar and GNSS satellites. A Trimble R10 was used as the base station with a Trimble R12 rover used to survey each check point. Additionally, the Trimble R10 acted as a GNSS base station for post-proccessing UAV trajectories.

The raw lidar data from May and August are converted to elevation maps using the LP360 software by GeoCue. The difference between the two elevation maps provides the ground truth elevation change. Our objective is to forecast the elevation change by using the lidar point cloud from May as input. Samples of point cloud, ground truth map, and forecasted map can be seen in Figure \ref{fig:qualitative}. 

%% file: Sections/4_machine_learning_methods.tex
\section{Methodology}

\subsection{Point Cloud Creation}
%The raw lidar files are processed using the LP360 software by GeoCue. 
The LP360 software offers a simple pipeline that steps through each stage of lidar point cloud creation. Importantly, UAV trajectories are corrected based on post-processed base station location and updated GNSS satellite orbits. Finally, images collected in-flight are matched across the point cloud resulting in a final point cloud with RGB color information. The final point cloud was checked for any misalignment in geolocation with ground control points. The processed point cloud was exported alongside an interpolated digital terrain model (DTM) based on the classified ground points of the point cloud. For each time step this gives us a full colorized and classified point cloud as well as an interpolated DTM for further use.

Given the colored point cloud $\mathcal{P} = \{(\mathbf{p}_i, \mathbf{f}_i)\}_{i=1}^{N}$ where $\mathbf{p}_i \in \mathbb{R}^3$ denotes spatial coordinates and $\mathbf{f}_i \in \mathbb{R}^4$ denotes per-point attributes (RGB, intensity), our goal is to predict a dense thaw depth map $\mathbf{Y} \in \mathbb{R}^{H \times W}$. We frame this as a 3D-to-2D projection problem analogous to bird's-eye-view perception in autonomous driving, but targeting continuous regression rather than object detection.

\begin{figure*}[t]
    \centering
    \includegraphics[width=\textwidth]{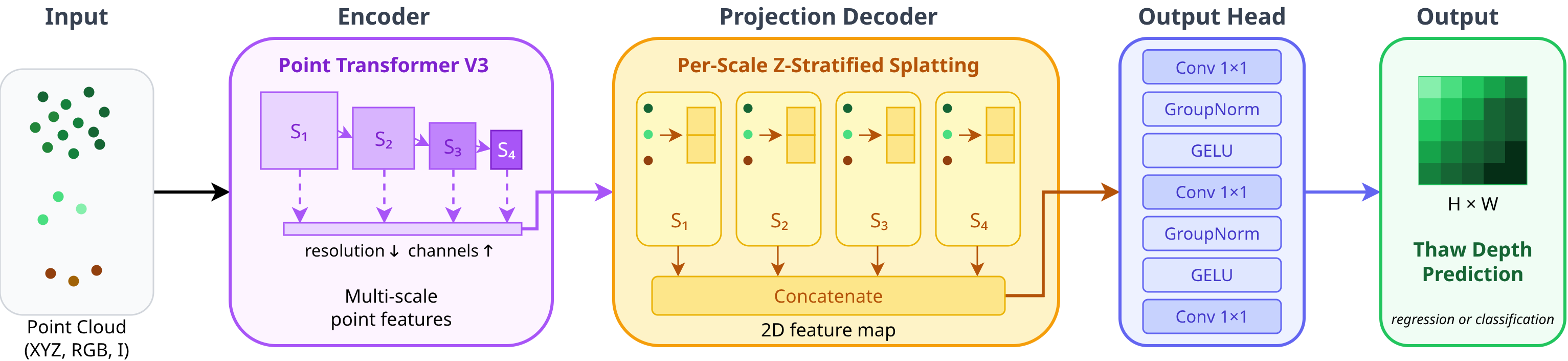}
    \caption{\textbf{Model architecture overview.} A point cloud with per-point features (XYZ, RGB, intensity) is processed by a Point Transformer V3 encoder, which produces multi-scale point features at four hierarchical stages. Each stage is independently projected to a 2D feature map via our height-aware projection mechanism with learned z-embeddings (see Fig.~\ref{fig:z_profile}), preserving vertical forest structure during the 3D-to-2D transformation. The resulting feature maps are concatenated and fused through $1\times1$ convolutions. A lightweight convolutional head produces the final per-pixel thaw depth prediction, supporting both regression and classification formulations.}
    \label{fig:architecture}
    \vspace{-3mm}
\end{figure*}

\subsection{Encoder}
We adopt Point Transformer V3 (PTv3)~\cite{PointTransformerV3} as our point cloud encoder. PTv3 processes 7-dimensional input (XYZ coordinates, RGB, and intensity) through a hierarchical architecture, producing multi-scale point features at four stages with channel dimensions $\{64, 128, 256, 512\}$. Each stage captures structure at a different spatial granularity: early stages preserve fine-grained local detail while later stages encode broader context. We leverage all four scales in our decoder, treating the encoder as a fixed architectural component and focusing our contribution on the projection mechanism. The entire architecture is shown graphically in Fig. \ref{fig:architecture}.

\subsection{Multi-Scale Z-Stratified Projection Decoder}

\begin{figure*}[t]
    \centering
    \includegraphics[width=\linewidth]{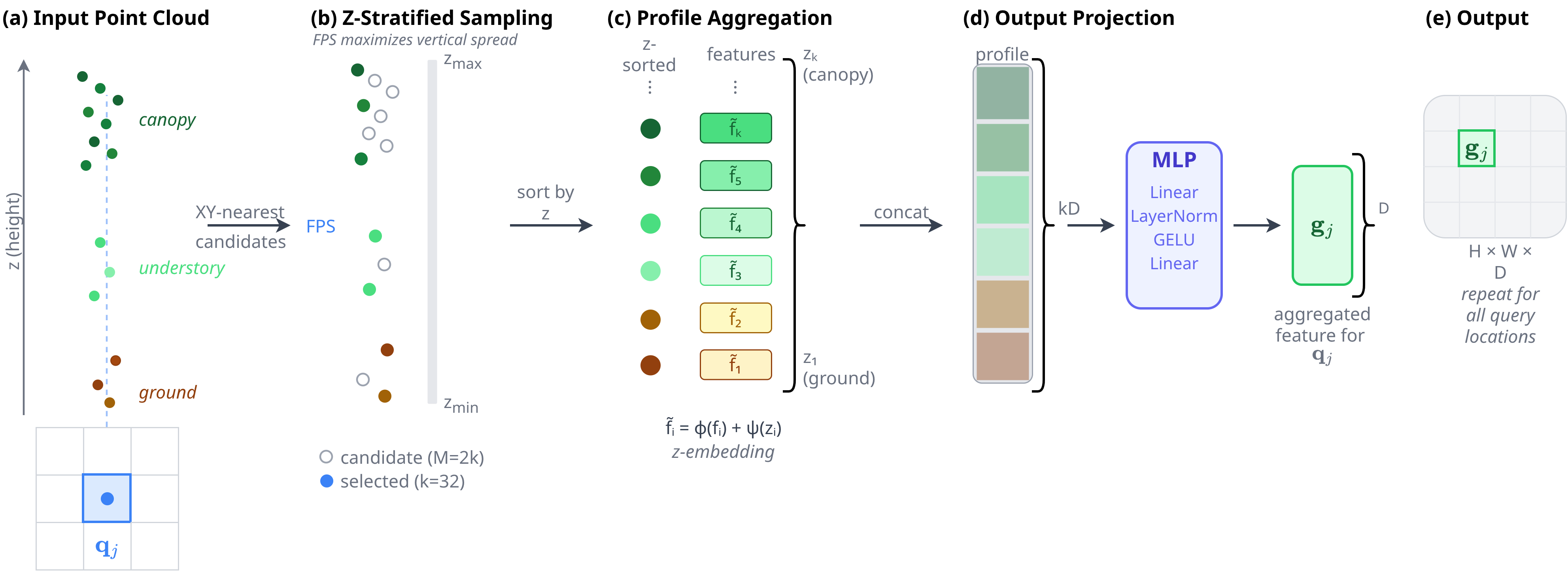}
    \caption{\textbf{Height-aware projection mechanism.} (a)~Input point cloud with vertical forest structure (ground, understory, canopy) above a query grid cell $q_j$. (b)~From XY-nearest candidates ($M = 2k$), farthest point sampling in the z-dimension selects $k$ points that span the full vertical extent. (c)~Selected points are sorted by height and augmented with learned z-embeddings that enable height-dependent feature transformations. (d)~The concatenated profile vector ($k \times D$) is projected through an MLP to produce a single aggregated feature. (e)~This process repeats for all query locations, yielding an $H \times W \times D$ feature map.}
    \label{fig:z_profile}
    \vspace{-4mm}
\end{figure*}

The core challenge in projecting 3D point features to a 2D grid lies in preserving height-dependent information. In forest environments, features from ground, understory, and canopy carry distinct signals about subsurface conditions, yet naive aggregation conflates them. We propose a decoder that (1) ensures all vertical strata are represented through stratified sampling, and (2) learns height-dependent feature transformations via explicit z-embeddings.

We propose a multi-scale decoder that processes each encoder stage independently before fusing their 2D representations. This late-fusion strategy allows fine scales to capture local geometric detail while coarse scales provide global context. This overall 3D to 2D projection module is shown graphically in Fig. \ref{fig:z_profile}.

For each encoder stage $s \in \{1, 2, 3, 4\}$, we first project point features to a common dimension $D = 128$ via a learned linear layer. We then define a regular grid of $H \times W = 64 \times 64$ query locations $\mathcal{Q} = \{\mathbf{q}_j\}_{j=1}^{HW}$ in normalized $[-1, 1]^2$ space, where each query represents a cell center for feature aggregation.

For each query $\mathbf{q}_j$, we identify $M = 2k$ candidate points nearest in XY-distance, where $k = 32$ is the target neighborhood size. Rather than selecting the $k$ nearest points, we apply farthest point sampling (FPS) over the $M$ candidates in the z-dimension to select $k$ points. FPS iteratively builds a subset by always selecting the point maximally distant from all previously selected points; applied to z-coordinates, this produces a set that spans the full vertical extent of the local point distribution. This ensures all forest strata remain represented regardless of point density.

Selected points are augmented with positional context. For each selected point $(\mathbf{p}_i,\mathbf{f}_i)$ in query $\mathbf{q}_j$'s neighborhood, we embed the z-coordinate in $\mathbf{p}_i$ through a learnable MLP $\psi: \mathbb{R} \rightarrow \mathbb{R}^D$ and add it to the projected features $\phi(\mathbf{f}_i)$:
\begin{equation}
    \tilde{\mathbf{f}}_i = \phi(\mathbf{f}_i) + \psi(z_i)
\end{equation}
where $\phi$ is the stage-specific linear projection. This embedding allows the network to learn height-dependent feature transformations.

We aggregate neighborhood features by exploiting vertical ordering. The $k$ selected points are sorted by z-coordinate (ground to canopy), and their features are concatenated into a single vector encoding the full vertical profile:
\begin{equation}
    \mathbf{v}_j = [w_1\tilde{\mathbf{f}}_{\pi(1)}; w_2\tilde{\mathbf{f}}_{\pi(2)}; \ldots; w_k\tilde{\mathbf{f}}_{\pi(k)}]
\end{equation}
where $\pi$ is the z-sorting permutation and $w_i$ are spatially-decaying weights based on XY-distance:
\begin{equation}
    w_i = \exp\bigl(-\lambda \cdot \max(0, \|\mathbf{q}_j - \mathbf{p}_i^{xy}\| - \tau)\bigr)
\end{equation}
with threshold $\tau = 0.1$ and falloff $\lambda = 10$. Points within the threshold receive unit weight; beyond it, influence decays exponentially. This is to prevent extremely far points from contributing excessively in sparse areas.

The concatenated profile $\mathbf{v}_j \in \mathbb{R}^{kD}$ is projected back to $\mathbb{R}^D$ via a two-layer MLP with LayerNorm and GELU activation, yielding the aggregated feature $\mathbf{g}_j$ for query location $j$.

Each scale produces an independent feature map $\mathbf{G}^{(s)}=[\mathbf{g}_j] \in \mathbb{R}^{H \times W \times D}$. We concatenate these along the channel dimension and fuse via $1 \times 1$ convolutions:
\begin{equation}
    \mathbf{Y} = \text{Conv}_{1\times1}\bigl([\mathbf{G}^{(1)}; \mathbf{G}^{(2)}; \mathbf{G}^{(3)}; \mathbf{G}^{(4)}]\bigr)
\end{equation}
with intermediate GroupNorm and GELU activations, progressively reducing from $4D$ channels to the final output dimension (1 for regression, $C$ for classification).

%% file: Sections/5_experiments.tex
\section{Experiments}
We evaluate our method on a dataset of aerial lidar collected over permafrost terrain in Fairbanks, Alaska. The dataset comprises temporally paired acquisitions: May imagery serves as model input, while August imagery provides ground truth after seasonal thaw. Ground truth thaw depth is derived from interpolated elevation change between the two acquisitions (see Section~\ref{sec:data_collection}) \footnote{The dataset and codes will be publicly available.}.

The continuous thaw depth raster is partitioned into $64 \times 64$ spatial tiles. Tiles containing data voids, or errors (tiles with change values greater than $\pm 1m$) are excluded, yielding 781 valid samples. From these valid samples, each fifth was designated as an evaluation tile, leaving 624 tiles for training and 157 for evaluation.

\vspace{-1mm}
\subsection{Preprocessing}

For each tile, we extract all points from the May acquisition whose XY coordinates fall within the tile boundary. Spatial X and Y coordinates are normalized to $[-1, 1]$ (tile-local), while Z is normalized to $[0, 1]$ globally across the dataset to preserve relative elevation information. Per-point features (RGB, intensity) are standardized to zero mean and unit variance. Ground truth thaw depths are transformed to approximate a standard normal distribution. We clip the 1st and 99th percentiles to remove outliers before normalization. Predictions are denormalized for evaluation.

\vspace{-1mm}
\subsection{Implementation Details}

We implement our model in PyTorch using Point Transformer V3 as the encoder backbone. We use $[2, 2, 2, 2]$ layers in each stage. The decoder uses hidden dimension $D = 128$, $k = 32$ neighbors, candidate multiplier $M = 2$, XY falloff threshold $\tau = 0.1$ and falloff rate $\lambda = 10$. More discussion on hyperparameter selection can be found in the Appendix. Output grid resolution is $64 \times 64$. PTv3 is trained from scratch on our dataset.

We use AdamW optimizer with learning rate $1\times10^{-5}$ and a polynomial learning scheduler. We start training with a 2 epoch warm up linearly ramping from 10\% to the full learning rate. The model is trained for a total of 100 epochs. We use batch size 1 with gradient accumulation over 2 steps, yielding an effective batch size of 2. Using Farthest Point Sampling (FPS), we cap the upper limit of points per tile to 60,000. Mixed precision (FP16) training and Flash Attention are also used to improve training speed and reduce memory size. Data augmentation includes random 90$^\circ$ rotations and random jitter in the xyz coordinates.
We evaluate after each epoch and retain the checkpoint with lowest validation loss. 

\vspace{-1mm}
\subsection{Evaluation Metrics}

Regression and classification models are trained with mean squared error (MSE) and cross entropy with inverse square weights. 
We report task-specific metrics on the held-out test set.

\textbf{Regression.}
For continuous thaw depth prediction, we report root mean squared error (RMSE), mean absolute error (MAE), and coefficient of determination ($R^2$). All error metrics are computed in original units (centimeters) after denormalization.

\textbf{Classification.}
While continuous predictions offer fine-grained detail, operational deployment often benefits from discrete severity categories. We evaluate a seven-class formulation spanning High Heave to High Thaw, with boundaries and class distributions shown in Table \ref{tab:class_distribution}. To measure classification performance, we report both class specific and mean Intersection over Union (mIoU). Additionally, since the classes form an ordinal scale from thaw to heave, we additionally report metrics that account for class ordering. 

\begin{table}[t]
    \centering
    \caption{Class distribution for dataset after discretizing the continuous ground truth labels.}
    \label{tab:class_distribution}
    \vspace{0.5em}
    \begin{tabular}{lcc}
        \toprule
        Class & Boundary (cm) & Percentage \\
        \midrule
        High Heave & $>1.6$ & 24.8\% \\
        Medium-high Heave & $[1.6, 1.0)$ & 16.2\% \\
        Medium-low Heave & $[1.0, 0.5)$ & 15.9\% \\
        Low Heave & $[0.5, 0.2)$ & 8.5\% \\
        No Change & $[0.2, -0.2]$ & 11.4\% \\
        Low Thaw & $(-0.2, -1.0]$ & 16.1\% \\
        High Thaw & $<-1.0$ & 7.1\% \\
        \bottomrule
    \end{tabular}
    \vspace{-3mm}
\end{table}

Quadratic Weighted Kappa (QWK) measures agreement between predictions and ground truth while penalizing disagreements quadratically by ordinal distance:
\begin{equation}
    \text{QWK} = 1 - \frac{\sum_{i,j} w_{ij} \, O_{ij}}{\sum_{i,j} w_{ij} \, E_{ij}}, \quad w_{ij} = \frac{(i - j)^2}{(K-1)^2}
\end{equation}
where $O_{ij}$ is the observed confusion matrix, $E_{ij}$ is the expected confusion matrix under random agreement, $K$ is the number of classes, and $w_{ij}$ weights disagreements by squared class distance. QWK ranges from $-1$ (systematic disagreement) to $1$ (perfect agreement), with $0$ indicating chance-level performance.

Mean Absolute Error in Class Units (MAECU) directly measures average ordinal displacement:
\begin{equation}
    \text{MAECU} = \frac{1}{N} \sum_{n=1}^{N} | \hat{y}_n - y_n |
\end{equation}
where $\hat{y}_n$ and $y_n$ are predicted and true class indices, respectively. Unlike mIoU, both metrics capture whether misclassifications fall to neighboring classes (minor error) or opposite ends of the scale (severe error).

\subsection{Baselines}
To evaluate the contribution of our proposed decoder, we compare it against a multi-scale mean-pooling baseline that shares the same PTv3 encoder and CNN refinement architecture. For each encoder stage $i$, point features $\mathbf{f}_j \in \mathbb{R}^{D_i}$
 are first projected to a common dimension via a learned linear layer. Points are then binned to a 2D grid based on their XY coordinates, with features aggregated via mean pooling:
\begin{equation}
    \mathbf{g}_{xy} = \frac{1}{|\mathcal{N}_{xy}|} \sum_{j \in \mathcal{N}_{xy}} \mathbf{\tilde{f}}_j
\end{equation}
where $\mathcal{N}_{xy}$ denotes the set of points falling within grid cell $(x, y)$, and $\mathbf{\tilde{f}}_j$ are features projected to the common dimension. The grids from all encoder stages are concatenated channel-wise and passed through the same convolutional refinement network used by our method. This baseline represents a straightforward application of PTv3 to dense prediction. It utilizes multi-scale fusion and identical output processing, differing only in the 3D-to-2D projection mechanism. Crucially, mean pooling across all heights causes ground-level information to be diluted by canopy returns, which dominate point density in forested scenes. Our z-stratified decoder addresses this by explicitly preserving vertical structure during projection.

To isolate whether categorical vegetation structure alone predicts thaw depth, we evaluate a histogram baseline that uses no learned point features. For each grid cell, we compute the proportion of points belonging to each LP360 classification category (ground, low vegetation, medium vegetation, high vegetation) along with log point density. This 6-channel feature map (5 class proportions + density) is processed by a lightweight CNN with three convolutional layers. This baseline tests whether the predictive signal is simply "what vegetation types are present" versus learned geometric and radiometric features.

\subsection{Results}
We evaluate our approach under both regression and classification formulations. Regression directly predicts continuous thaw depth, while classification assigns pixels to discrete severity categories.

\paragraph{Regression.}
Table~\ref{tab:regression} presents regression performance on the held-out evaluation set. The histogram approach performs worst (R$^2=0.656$), confirming that categorical vegetation proportions alone poorly predict thaw depth. Mean-pooling improves (R$^2=0.705$) by leveraging learned point features, but conflates  canopy and ground-level information. Our decoder substantially outperforms both, reducing RMSE by $70\%$ relative to mean pooling ($0.515 \rightarrow 0.121 cm$) and achieving R$^2=0.984$. These gains demonstrate that height-dependent feature encoding is vital for accurate thaw prediction. 

\begin{table}[t]
    \centering
    \caption{Regression results. Best results in \textbf{bold}.}
    \label{tab:regression}
    \vspace{0.5em}
    \begin{tabular}{lccc}
        \toprule
        Method & RMSE $\downarrow$ & MAE $\downarrow$ & $R^2$ $\uparrow$ \\
        \midrule
        Histogram & 0.556 & 0.460 & 0.656 \\
        Mean-pooling & 0.515 & 0.265 & 0.705 \\
        \textbf{Ours} & \textbf{0.121} & \textbf{0.089} & \textbf{0.984} \\
        \bottomrule
    \end{tabular}
    \vspace{-5mm}
\end{table}

\begin{table*}[t]
    \centering
    \caption{Classification results. Per-class IoU and mean IoU (\%). QWK = Quadratic Weighted Kappa; MAECU = Mean Absolute Error in Class Units.}
    \label{tab:classification}
    \vspace{0.5em}
    \begin{tabular}{lccccccc|ccc}
        \toprule
        Method & C1 & C2 & C3 & C4 & C5  & C6 & C7 & mIoU $\uparrow$ & QWK $\uparrow$ & MAECU $\downarrow$ \\
        \midrule
        Histogram & 50.2 & 21.8 & 14.3 & 11.9 & 11.1 & 29.0 & 24.6 & 23.3 & 0.52 & 1.27 \\
        
        Mean-pooling & 70.8 & \textbf{67.7} & 65.1 & 42.8 & 54.7 & 63.5 & 37.9 & 57.5 & 0.75 & 0.53 \\
        
        \textbf{Ours} & \textbf{89.4} & 67.5 & \textbf{68.0} & \textbf{43.3} & \textbf{60.1} & \textbf{67.6} & \textbf{65.8} & \textbf{66.1} & \textbf{0.97} & \textbf{0.19} \\
        \bottomrule
    \end{tabular}
\end{table*}

\paragraph{Classification.}
Table~\ref{tab:classification} reports classification performance. The histogram baseline achieves only 23.3 mIoU, demonstrating that categorical vegetation proportions alone are insufficient for thaw prediction despite vegetation structure's known correlation with permafrost dynamics. The mean-pooling baseline improves substantially to 57.5 mIoU, indicating that learned point features capture information beyond vegetation categories. Our method achieves a $8.6$-point improvement in mIoU over the mean-pooling baseline, with the largest gains at the ordinal extremes: $+18.6$ points for High heave (C1) and $+27.9$ points for High Thaw (C7). Accurate discrimination at the extremes is particularly important for infrastructure monitoring, where confusing severe thaw with heave could lead to opposite intervention decisions. 
The ordinal metrics reveal more about how the models misclassify. QWK improves from $0.52$ (histogram) to $0.75$ (mean-pooling) to $0.97$ (ours), indicating that when our model does misclassify, predictions remain closer to the true severity on the thaw-heave spectrum. This is corroborated by the reduction in MAECU from $0.53$ (mean-pooling) to $0.19$ class units. This shows that explicitly augmenting features by height enables the network to better distinguish ground surface conditions. 

\begin{figure*}[t]
    \centering
    \includegraphics[width=\linewidth]{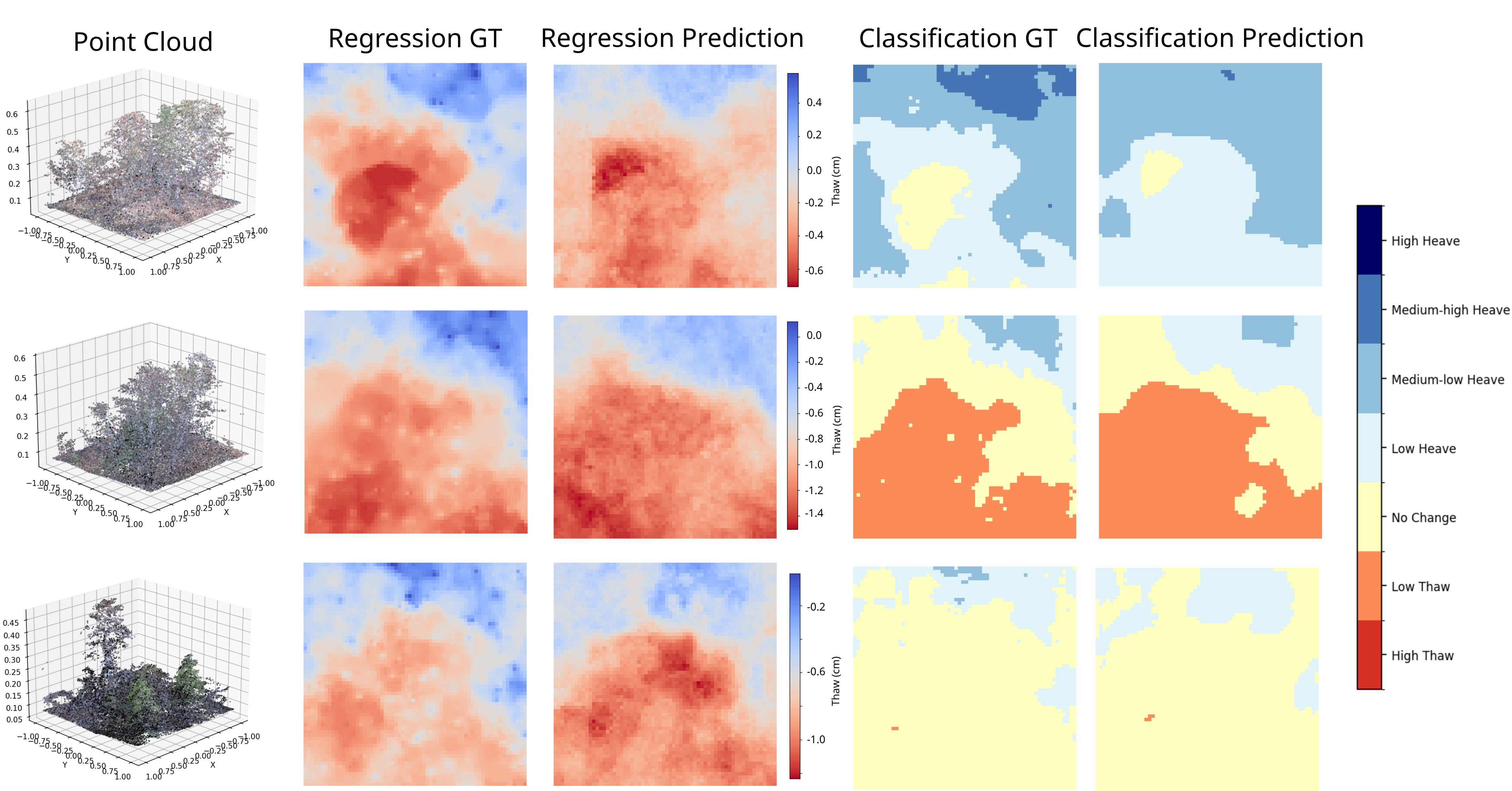}
    \caption{Qualitative comparison. Each row: input point cloud, regression ground truth, regression prediction, classification ground truth, classification prediction.}
    \label{fig:qualitative}
    \vspace{-3mm}
\end{figure*}

\subsection{Ablation Study}

We ablate key decoder components in Table \ref{tab:ablation}, showing both classification and regression results.

\begin{table}[t]
\centering
\caption{Ablation study on decoder components.}
\label{tab:ablation}
    \vspace{0.5em}
\begin{tabular}{lcc}
    \toprule
    Variant & mIoU (\%)  & RMSE (cm) \\
    \midrule
    Full model & 66.1 & 0.121 \\
    \midrule
    \multicolumn{2}{l}{\textit{Projection strategy}} \\
    \quad Mean pooling (no profile) & 62.3 & 0.297\\
    \quad Closest-k (no FPS in z) & 65.3 & 0.151\\
    \quad $M=4$ candidates & 62.9 & 0.272 \\
    \midrule
    \multicolumn{2}{l}{\textit{Feature encoding}} \\
    \quad w/o z-embedding & 58.7 & 0.448\\
    \midrule
    \multicolumn{2}{l}{\textit{Multi-scale fusion}} \\
    \quad No multi-scale ($S_4$ only)  & 61.5 & 0.192\\
    \bottomrule
\end{tabular}
\vspace{-3mm}
\end{table}

Removing the learned z-embedding produces the largest performance drop ($-7.4$ points in mIoU), indicating that height-dependent feature modulation is the most critical component of our decoder. Without explicit elevation encoding, the model cannot distinguish features originating from canopy versus ground returns during aggregation, even when the projection strategy preserves their spatial separation. 

Replacing our stratified profile sampling with mean pooling reduces mIoU by $3.8$ points.  Mean pooling allows dense canopy returns to dominate the aggregated features, diluting the sparse but informative ground-level points that carry the strongest signal for thaw prediction. Taking only the closest $k$ neighbors, rather than using FPS to maximize the z-profile of each pixel, produces a drop in mIoU of $0.8$ points. This drop indicates that spreading out sampled points maximizes the information available to the decoder. Without it, sampled points may be extremely close together, which intuitively should carry redundant information if they represent the same object. Conversely, setting $M=4$ reduces miou by $3.2$ points. This is because a higher $M$ allows for farther points to be sampled by FPS during decoding, which are too far away to contribute meaningful local information. Finally, using only the final encoder stage ($S_4$) decreases mIoU by $4.6$ points. Earlier stages capture fine-grained local geometry at higher point density, while later stages provide broader semantic context. Fusing across scales allows the decoder to leverage both detailed surface structure and coarse vegetation patterns.

%% file: Sections/6_discussion.tex
\section{Discussion}
Our results demonstrate that learned height encoding is essential for accurate dense prediction in forested lidar scenes. The $70\%$ reduction in RMSE and near-perfect $R^2$ of $0.974$ indicate that our decoder successfully preserves ground-level information beneath dense canopy coverage, enabling larger gains at the extremes of the thaw-heave spectrum where accurate prediction matters most for downstream decision-making.

The histogram baseline's poor performance ($23.3$ mIoU) confirms that categorical vegetation structure alone is insufficient despite vegetation's established relationship with permafrost thermal dynamics. The 34-point gap between histogram and mean-pooling baselines indicates that PTv3 captures meaningful representations beyond categorical structure. The ablation study shows that learned z-embeddings contribute more to performance than the sampling strategy itself ($-7.4$ vs $-0.8$ mIoU). This suggests the decoder's effectiveness lies not in only including points from multiple heights, but in learning how to weight features based on their vertical origin, essentially functioning as a learned vertical attention mechanism. 

From an application perspective, forecasting sub-centimeter thaw depth from single-timepoint winter lidar enables proactive infrastructure management by identifying at-risk areas before thaw occurs The strong performance on extreme classes is the most valuable in this context. Distinguishing severe thaw from heave represents the difference between subsidence risk requiring intervention and frost-driven uplift that may self-correct.

This work is not without limitations. Most significantly, our evaluation is constrained to a single geographic site in interior Alaska, comprising 781 spatial tiles. While the held-out evaluation set was spatially disjoint from training data, the model has not been tested on different forest types, permafrost regimes, or climatic conditions. The extent to which learned representations transfer across sites remains an open question. 

Beyond permafrost monitoring, our approach addresses a general challenge in 3D scene understanding: projecting volumetric features to 2D when informative content is occluded by dominant structures. Applications include understory estimation in forestry, road surface prediction beneath vegetation, and ground-level inference in urban scenes. Future work should prioritize cross-site validation and multi-temporal sequences to learn seasonal dynamics. Interpreting the learned z-embeddings' relationship to vegetation strata could yield insights for both remote sensing and permafrost modeling.

%% file: Sections/7_conclusion.tex
\section{Conclusion}
We presented a projection decoder with learned height embedding for dense prediction from aerial lidar in forested environments. By augmenting point features with explicit z-encodings, our approach enables the model to learn height-dependent feature transformations that differentiate ground-level signals from canopy returns. Combined with stratified sampling that ensures all forest strata remain represented, the vertical information critical for predicting subsurface conditions is preserved. This enables thaw depth prediction from single-date winter point clouds, achieving a $70\%$ reduction in RMSE and near-perfect ordinal agreement (QWK = $0.97$) on our permafrost monitoring task.

The ablation analysis reveals that learned z-embeddings contribute more to performance than the projection strategy itself, suggesting that the decoder's effectiveness lies in learning height-dependent feature transformations rather than just sampling across strata. This finding has many implications beyond permafrost: any dense prediction task where vertically-distributed 3D structure must be collapsed to 2D (understory estimation in forestry, road surface prediction beneath vegetation, ground-level inference in occluded urban scenes) may benefit from explicit elevation encoding during projection.

%Our evaluation is limited to a single site in interior Alaska, and the extent to which learned representations generalize across permafrost regimes, forest types, and sensor configurations remains to be established. Validating transferability across diverse Arctic and sub-Arctic environments is a priority for future work, alongside investigation of what physical or ecological properties the learned z-embeddings capture. More broadly, we hope this work demonstrates the value of designing projection mechanisms that respect the structure of environmental data, rather than treating 3D-to-2D conversion as a solved preprocessing step.

% \section*{Acknowledgements}
% \textbf{Do not} include acknowledgements in the initial version of the paper

% If a paper is accepted, the final camera-ready version can (and usually should)
% include acknowledgements.  Such acknowledgements should be placed at the end of
% the section, in an unnumbered section that does not count towards the paper
% page limit. Typically, this will include thanks to reviewers who gave useful
% comments, to colleagues who contributed to the ideas, and to funding agencies
% and corporate sponsors that provided financial support.
\section*{Impact Statement}
We present a projection decoder with learned height embeddings that enables accurate dense predictions from aerial lidar in forested environments, achieving sub-centimeter permafrost thaw estimation from single-date acquisitions. Our approach addresses the general challenge in 3D scene understanding of inferring ground-level properties beneath occluding structures, with immediate applications in Arctic infrastructure monitoring and climate change mitigation.

% Authors are \textbf{required} to include a statement of the potential broader
% impact of their work, including its ethical aspects and future societal
% consequences. This statement should be in an unnumbered section at the end of
% the paper (co-located with Acknowledgements -- the two may appear in either
% order, but both must be before References), and does not count toward the paper
% page limit. In many cases, where the ethical impacts and expected societal
% implications are those that are well established when advancing the field of
% Machine Learning, substantial discussion is not required, and a simple
% statement such as the following will suffice:

% ``This paper presents work whose goal is to advance the field of Machine
% Learning. There are many potential societal consequences of our work, none
% which we feel must be specifically highlighted here.''

% The above statement can be used verbatim in such cases, but we encourage
% authors to think about whether there is content which does warrant further
% discussion, as this statement will be apparent if the paper is later flagged
% for ethics review.

%% file: Sections/8_appendix.tex
\newpage
\appendix
\onecolumn

\section{Dataset Details}
\label{app:dataset}

\subsection{Ground Truth Raster Properties}

The ground truth elevation change raster was derived from temporally paired UAV-lidar acquisitions (May to August 2024), computed as (August elevation $-$ May elevation). Negative values indicate thaw (ground subsidence), while positive values indicate heave (frost uplift or vegetation growth). Table~\ref{tab:raster_props} summarizes the raster properties.

\begin{table}[h]
\centering
\caption{Ground truth raster properties.}
\label{tab:raster_props}
\begin{tabular}{ll}
\toprule
Property & Value \\
\midrule
Coordinate Reference System & EPSG:32606 (UTM Zone 6N) \\
Spatial Resolution & 0.10 m $\times$ 0.10 m \\
Raster Dimensions & 2261 $\times$ 2307 pixels \\
Geographic Extent & 226.1 m $\times$ 230.7 m \\
Total Area & 5.22 ha \\
Valid Data Coverage & 67.0\% \\
Valid Value Range & $[-2.33, +5.70]$ cm \\
\bottomrule
\end{tabular}
\end{table}

\subsection{Elevation Change Statistics}

Table~\ref{tab:gt_stats} presents comprehensive statistics for the ground truth elevation change distribution. The positive mean (+0.99 cm) and median (+0.72 cm) indicate the site is heave-dominated overall, with localized thaw features corresponding to active thermokarst (including a central pond). The positive skewness (+0.96) reflects the asymmetric distribution with a longer tail toward extreme heave values, including vegetation growth in a grassy strip along the eastern edge.

\begin{table}[h]
\centering
\caption{Ground truth elevation change statistics. Negative values indicate thaw (subsidence); positive values indicate heave (uplift).}
\label{tab:gt_stats}
\begin{tabular}{lrlr}
\toprule
Statistic & Value (cm) & Statistic & Value \\
\midrule
Minimum (max thaw) & $-2.33$ & Skewness & $+0.96$ \\
Maximum (max heave) & $+5.70$ & Kurtosis & $0.45$ \\
Mean & $+0.99$ & Coef. of Variation & $164.6\%$ \\
Median & $+0.72$ & Valid Pixels & 3,495,868 \\
Std. Dev. & $1.62$ & \\
\bottomrule
\end{tabular}
\end{table}

\subsection{Thaw vs.\ Heave Distribution}

\begin{table}[h]
\centering
\caption{Overall distribution of thaw and heave pixels.}
\label{tab:thaw_heave}
\begin{tabular}{lrr}
\toprule
Category & Pixels & Percentage \\
\midrule
Thaw (negative, subsidence) & 808,730 & 23.1\% \\
No Change ($\pm$0.2 cm) & 396,961 & 11.4\% \\
Heave (positive, uplift) & 2,290,177 & 65.5\% \\
\bottomrule
\end{tabular}
\end{table}

\subsection{Spatial Autocorrelation}

We quantify spatial autocorrelation using Moran's I statistic, which measures the degree to which values at nearby locations are more similar (positive autocorrelation) or more dissimilar (negative autocorrelation) than expected under spatial randomness. Moran's I is defined as:

\begin{equation}
I = \frac{N}{\sum_{i}\sum_{j} w_{ij}} \cdot \frac{\sum_{i}\sum_{j} w_{ij}(x_i - \bar{x})(x_j - \bar{x})}{\sum_{i}(x_i - \bar{x})^2}
\label{eq:morans_i}
\end{equation}

where $N$ is the number of spatial units (pixels), $x_i$ is the value at location $i$, $\bar{x}$ is the global mean, and $w_{ij}$ is the spatial weight between locations $i$ and $j$ (typically 1 for adjacent pixels and 0 otherwise). The statistic ranges from $-1$ (perfect dispersion) through $0$ (spatial randomness) to $+1$ (perfect clustering).

The ground truth exhibits strong positive spatial autocorrelation (Moran's I = 0.99), indicating that thaw and heave patterns form spatially coherent clusters rather than random noise. This high value confirms that elevation change is governed by spatially continuous physical processes (e.g., thermokarst expansion, frost heave) rather than pixel-independent noise. From a modeling perspective, this supports the use of convolutional operations in the output head, as neighboring predictions are highly correlated and spatial context provides meaningful information.

\section{Class Distribution Analysis}
\label{app:class}

\subsection{Per-Class Statistics}

Table~\ref{tab:class_stats} provides detailed statistics for each severity class. The dataset exhibits moderate class imbalance: High Thaw (C7) contains only 7.1\% of pixels, while High Heave (C1) contains 24.8\%---a 3.5$\times$ imbalance ratio. This reflects the physical reality that severe thermokarst subsidence is spatially localized (e.g., the pond feature), while heave dominates the surrounding stable permafrost and vegetated areas.

\begin{table}[h]
\centering
\caption{Per-class statistics for the 7-class formulation. Boundaries defined on elevation change (negative = thaw, positive = heave). Imbalance ratio computed as (max class count) / (class count).}
\label{tab:class_stats}
\begin{tabular}{llrrrr}
\toprule
Class & Boundary (cm) & Count & \% & Weight & Imbalance \\
\midrule
C1: High Heave & $> +1.6$ & 868,553 & 24.8 & 1.00 & 1.0$\times$ \\
C2: Med-High Heave & $(+1.0, +1.6]$ & 566,837 & 16.2 & 1.53 & 1.5$\times$ \\
C3: Med-Low Heave & $(+0.5, +1.0]$ & 556,133 & 15.9 & 1.56 & 1.6$\times$ \\
C4: Low Heave & $(+0.2, +0.5]$ & 298,654 & 8.5 & 2.91 & 2.9$\times$ \\
C5: No Change & $[-0.2, +0.2]$ & 396,961 & 11.4 & 2.19 & 2.2$\times$ \\
C6: Low Thaw & $[-1.0, -0.2)$ & 561,384 & 16.1 & 1.55 & 1.5$\times$ \\
C7: High Thaw & $< -1.0$ & 247,346 & 7.1 & 3.51 & 3.5$\times$ \\
\bottomrule
\end{tabular}
\end{table}

The heave-dominated distribution (65.5\% of pixels in C1--C4) reflects the Farmer's Loop site characteristics: predominantly stable permafrost with seasonal frost heave, a grassy strip exhibiting vegetation growth between acquisitions (+5.70 cm maximum), and localized thermokarst features including a central pond ($-2.33$ cm maximum subsidence). The 3.5$\times$ imbalance for High Thaw (C7) motivates our use of inverse-frequency class weighting and highlights the challenge of detecting rare but critical subsidence events for infrastructure monitoring.

\section{Implementation Details}
\label{app:implementation}

\subsection{Full Hyperparameter Configuration}
The complete set of hyperparameters are shown in Table \ref{tab:hyperparams}.
\begin{table}[h]
\centering
\caption{Complete hyperparameter configuration.}
\label{tab:hyperparams}
\begin{tabular}{lll}
\toprule
Component & Parameter & Value \\
\midrule
\textbf{Encoder (PTv3)} & Stages & 4 \\
& Layers per stage & [2, 2, 2, 2] \\
& Channel dimensions & [64, 128, 256, 512] \\
& Input features & 7 (XYZ, RGB, intensity) \\
\midrule
\textbf{Decoder} & Hidden dimension $D$ & 128 \\
& Neighbors $k$ & 32 \\
& Candidate multiplier $M$ & 2 \\
& Output grid size & $64 \times 64$ \\
& XY falloff threshold $\tau$ & 0.1 \\
& XY falloff rate $\lambda$ & 10 \\
\midrule
\textbf{Training} & Optimizer & AdamW \\
& Learning rate & $1 \times 10^{-5}$ \\
& LR scheduler & Polynomial decay \\
& Warmup epochs & 2 \\
& Total epochs & 100 \\
& Batch size & 1 \\
& Gradient accumulation & 2 steps \\
& Max points per tile & 60,000 \\
& Precision & Mixed (FP16) \\
\midrule
\textbf{Augmentation} & Rotation & Random 90° multiples \\
& Jitter & XYZ coordinate noise \\
\midrule
\textbf{Loss (Classification)} & Type & Cross-entropy \\
& Class weights & Inverse frequency \\
\midrule
\textbf{Loss (Regression)} & Type & MSE \\
\bottomrule
\end{tabular}
\end{table}

\subsection{XY Distance Weighting Parameters}
\label{app:xy_weighting}

The z-stratified projection decoder uses a threshold-based exponential weighting scheme to control how points contribute to each grid cell based on their XY distance. For each grid cell query location, the weight assigned to a neighboring point at XY distance $d$ is:
\begin{equation}
    w(d) = \exp\left(-\lambda \cdot \max(0, d - \tau)\right)
\end{equation}
where $\tau$ is the distance threshold below which points receive full weight, and $\lambda$ controls the decay rate beyond the threshold.

\begin{figure}[h]
\centering
\includegraphics[width=\textwidth]{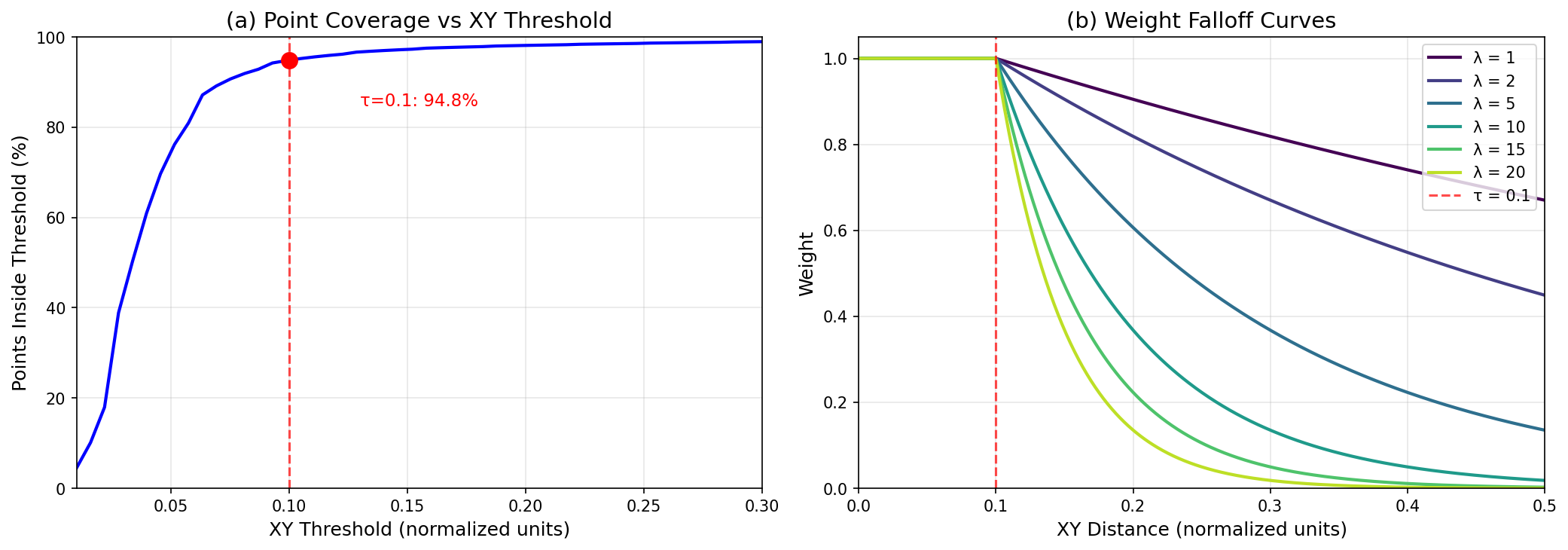}
\caption{XY distance weighting analysis. \textbf{Left:} Fraction of k-nearest neighbors within threshold distance. \textbf{Right:} Weight falloff curves for different $\lambda$ values at $\tau=0.1$.}
    \label{fig:xy_weighting}
\end{figure}

We select $\tau = 0.1$ and $\lambda = 10$ based on empirical analysis of our dataset's point distribution. Figure~\ref{fig:xy_weighting} shows the relationship between these parameters and point coverage.

\textbf{Threshold Selection ($\tau = 0.1$).} 
The left panel shows the percentage of k-nearest neighbors (k=64) that fall within a given threshold distance. We plot this number because when $M=2$, FPS is sampling from 64 of the nearest points to generate the Z-profile for the decoder. At $\tau = 0.1$, approximately 94.8\% of selected neighbors receive full weight ($w=1$). This ensures that the vast majority of points used in aggregation contribute without attenuation, while still allowing the weighting scheme to downweight the small fraction of points that may be spatially misaligned due to variance in point cloud density. A smaller threshold would unnecessarily penalize well-positioned points, while a larger threshold would effectively disable the weighting mechanism entirely.

\textbf{Falloff Selection ($\lambda = 10$).}
The right panel illustrates weight decay curves for various $\lambda$ values. With $\lambda = 10$, points at distance $d = 0.2$ (twice the threshold) receive weight $w \approx 0.37$, and points at $d = 0.3$ receive $w \approx 0.14$. This provides a smooth transition that maintains spatial locality without introducing discontinuities. Lower values (e.g., $\lambda = 1$) would allow distant points to contribute nearly equally, potentially blurring spatial boundaries. Higher values (e.g., $\lambda = 20$) would create sharper cutoffs that approach hard thresholding, which we found less stable during training.

The combination of $\tau = 0.1$ and $\lambda = 10$ thus implements a ``soft locality'' constraint: points within approximately one grid cell spacing contribute fully, while more distant points are progressively downweighted rather than excluded entirely.

\subsection{Neighborhood Size Selection}
\label{app:k_neighbors}

We use $k=32$ neighbors per grid cell, informed by both empirical analysis and established conventions in point cloud deep learning~\cite{PointNet++,PointTransformer}.

As described in Section~\ref{app:candidate_multiplier}, we first retrieve $M \cdot k = 64$ candidate points (the nearest neighbors in XY), then select $k=32$ from this pool. Crucially, farthest point sampling (FPS) in the z-dimension will preferentially select points near the spatial boundary of the candidate pool, since XY distance is not taken into consideration. This means the effective spatial extent of the final $k$ points is largely determined by the $M \cdot k$ candidate pool, not $k$ alone.

\begin{figure}[t]
    \centering
    \includegraphics[width=0.6\textwidth]{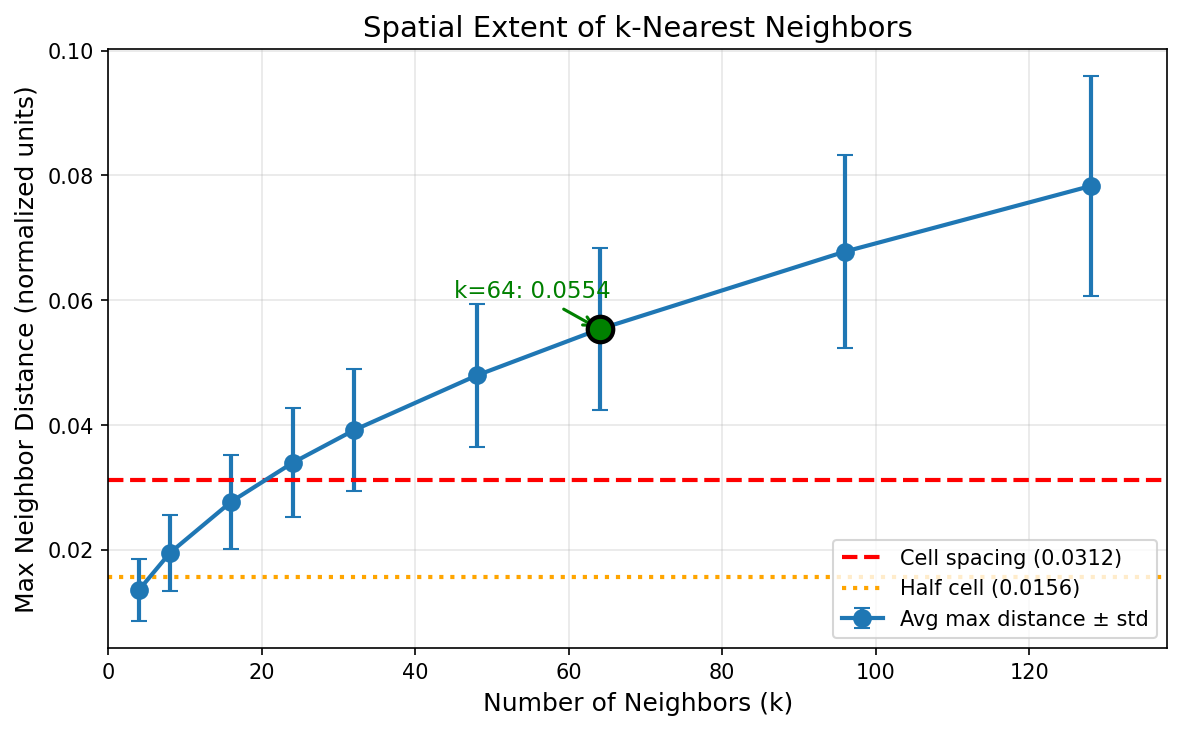}
    \caption{Spatial extent of k-nearest neighbors. The y-axis shows the average maximum distance among the k neighbors selected for each grid cell. The red dashed line indicates the grid cell spacing; values below this line indicate insufficient coverage.}
    \label{fig:k_coverage}
\end{figure}

Figure~\ref{fig:k_coverage} shows the average maximum distance among $k$-nearest neighbors as a function of $k$, computed across sample tiles from our dataset. The dashed line indicates the grid cell spacing in normalized coordinates ($2/64 \approx 0.031$). With $M=2$, our 64-point candidate pool extends well beyond the cell boundary, and FPS will reliably select points from this outer region to capture vertical extremes. This ensures complete spatial coverage while the XY weighting scheme (Section~\ref{app:xy_weighting}) appropriately downweights these more distant points during feature aggregation.

\subsection{Candidate Pool Multiplier}
\label{app:candidate_multiplier}

The hybrid sampling strategy first identifies $M \cdot k$ candidate points (the $M \cdot k$ nearest neighbors in XY), then selects $k$ points from this pool using a combination of closest-k and farthest point sampling (FPS) in the z-dimension. The multiplier $M$ controls the size of this candidate pool.

Larger $M$ provides more candidates for FPS to select from, potentially enabling better vertical diversity. However, expanding the candidate pool necessarily includes points at greater XY distances, which may be spatially misaligned with the target grid cell. This creates a tradeoff between vertical sampling diversity and horizontal spatial fidelity.

At $M=1$, the candidate pool equals the final selection size ($M \cdot k = k$), reducing the method to simple k-nearest neighbors. This eliminates the z-stratified sampling that motivates our decoder design.
We evaluated $M \in \{2, 4\}$ and found that $M=2$ achieves 66.1 mIoU compared to 62.9 mIoU for $M=4$. This 3.1 point improvement demonstrates that the candidate pool should be kept spatially tight. With $M=4$ (128 candidates for $k=32$), the outer candidates lie at XY distances where even the exponential downweighting cannot fully compensate for spatial misalignment. The resulting feature aggregation conflates information from neighboring grid cells, degrading prediction accuracy. In contrast, $M=2$ (64 candidates) restricts the pool to a compact spatial neighborhood while still providing sufficient candidates for FPS to achieve meaningful vertical diversification.

\subsection{Data Normalization}

\begin{itemize}
    \item \textbf{Spatial coordinates:} X, Y normalized to $[-1, 1]$ per-tile; Z normalized to $[0, 1]$ globally across the dataset to preserve relative elevation information.
    \item \textbf{Per-point features:} RGB and intensity standardized to zero mean, unit variance using dataset-wide statistics.
    \item \textbf{Ground truth:} Normalized to zero mean and unit variance for training. Predictions are denormalized for evaluation.
\end{itemize}

\section{Computational Cost}
\label{app:compute}

\begin{table}[h]
\centering
\caption{Computational requirements. Inference time measured on an NVIDIA RTX 4090M Laptop GPU.}
\label{tab:compute}
\begin{tabular}{lccc}
\toprule
Method & Parameters & GPU Memory & Inference \\
\midrule
Histogram Baseline & 0.2M & 0.02 GB & 0.76 ms/tile \\
Mean-pooling + PTv3 & 39.4M & 4.3 GB & 119 ms/tile \\
Ours (Z-Stratified) & 41.1M & 4.6 GB & 133 ms/tile \\
\bottomrule
\end{tabular}
\end{table}

Our decoder adds 1.7M parameters (+4.3\%) and 14ms inference time (+11.8\%) compared to mean-pooling, while achieving 8.6-point mIoU improvement and 70\% RMSE reduction---a favorable accuracy-efficiency tradeoff.

\section{Broader Applicability}
\label{app:broader}

While developed for permafrost monitoring, our z-stratified projection addresses a general challenge: preserving vertically-distributed information when projecting 3D data to 2D beneath occluding structures. Potential applications include:

\begin{itemize}
    \item \textbf{Forestry:} Understory biomass estimation from canopy-penetrating lidar
    \item \textbf{Urban mapping:} Ground-level inference beneath building overhangs
    \item \textbf{Autonomous driving:} Road surface estimation under vegetation
    \item \textbf{Archaeology:} Subsurface feature detection through forest cover
\end{itemize}

\section{Ground Truth Visualization and Analysis}
\label{app:visualization}

This section presents visualizations of the ground truth elevation change data to provide insight into the dataset characteristics and support reproducibility.

\subsection{Value Distribution}

\begin{figure}[h]
\centering
\includegraphics[width=\textwidth]{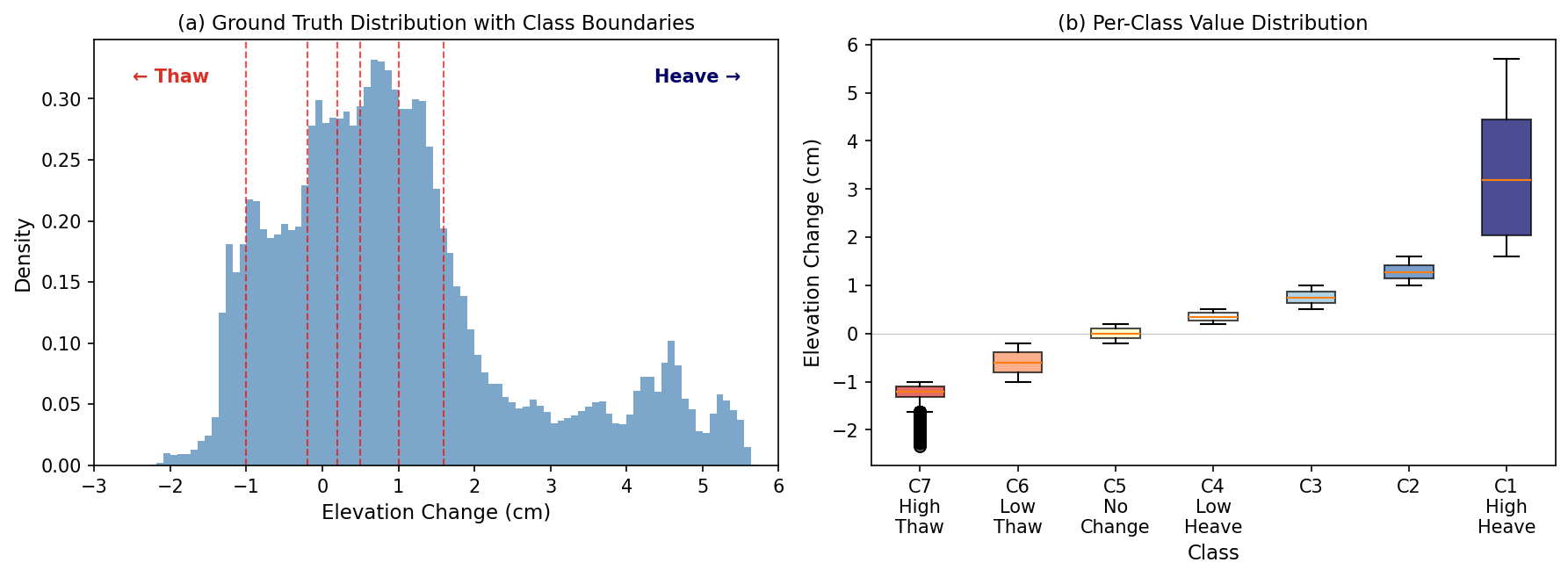}
\caption{Ground truth value distribution. (a) Histogram of elevation change values with class boundaries (dashed red lines). Negative values indicate thaw (subsidence); positive values indicate heave (uplift). (b) Box plots showing the value distribution within each class, ordered from High Thaw (C7) to High Heave (C1).}
\label{fig:histogram}
\end{figure}

Figure~\ref{fig:histogram} shows the distribution of elevation change values across the study site. The histogram (a) reveals a right-skewed distribution centered around $+0.7$ cm, with the majority of pixels falling in the heave regime (positive values). The vertical dashed lines indicate the class boundaries used for the 7-class classification formulation. The distribution exhibits a clear peak in the low heave range ($0$ to $+1.0$ cm) and a longer tail toward extreme heave values reaching $+5.7$ cm, corresponding to vegetation growth in grassy areas.

The box plots (b) illustrate the within-class value distributions, ordered from High Thaw (C7, leftmost) to High Heave (C1, rightmost). The thaw classes (C6--C7) show tight distributions corresponding to the localized thermokarst feature, while High Heave (C1) exhibits greater spread due to varying vegetation growth rates. The median values progress monotonically across classes, confirming that the ordinal class structure captures a meaningful physical gradient from subsidence to uplift.

\subsection{Spatial Distribution}

\begin{figure}[h]
\centering
\includegraphics[width=\textwidth]{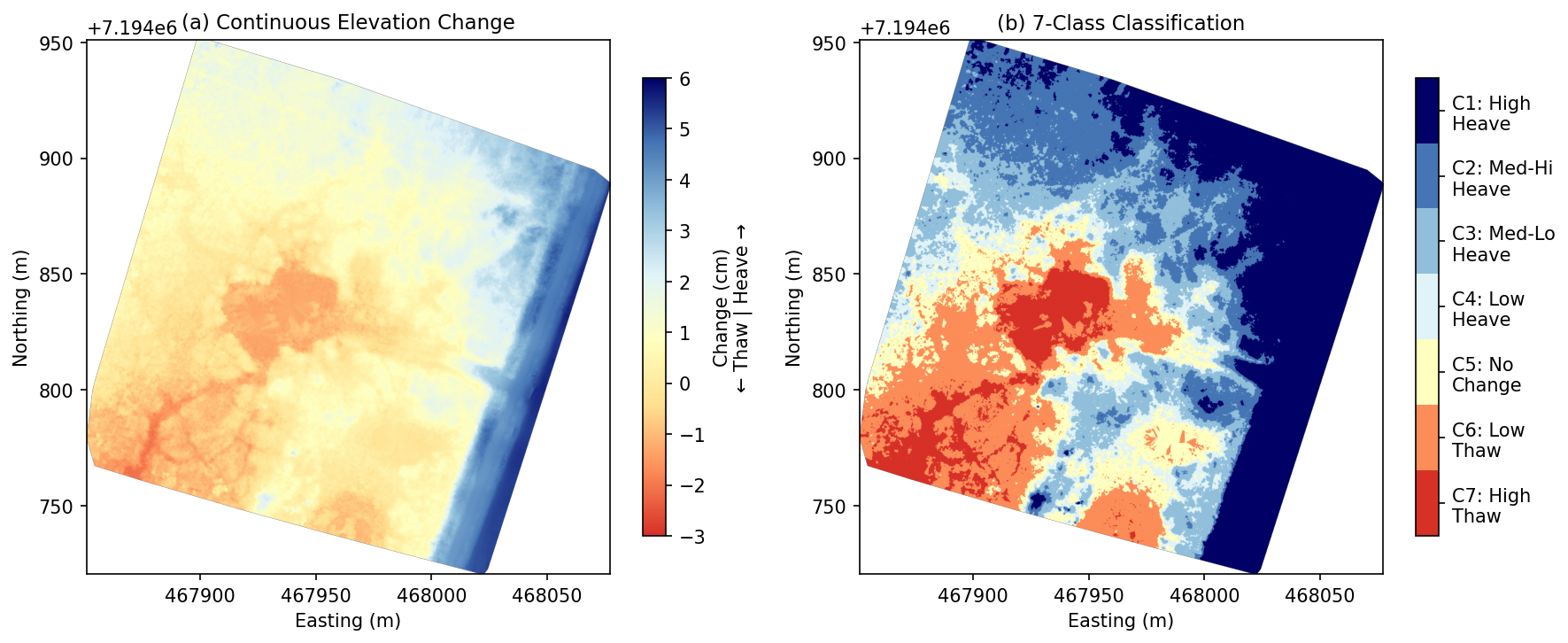}
\caption{Spatial distribution of ground truth elevation change. (a) Continuous values in centimeters, with warm colors indicating thaw (subsidence) and cool colors indicating heave (uplift). (b) 7-class discretization showing the spatial extent of each severity category. Coordinates are in UTM Zone 6N (EPSG:32606).}
\label{fig:spatial}
\end{figure}

Figure~\ref{fig:spatial} presents the spatial distribution of ground truth values across the Farmer's Loop study site. The continuous map (a) reveals distinct spatial patterns: a localized thaw feature (thermokarst with central pond) appears in warm colors (red/orange) in the south-central region, while heave dominates the surrounding areas (cool colors), with the most extreme heave along a grassy strip on the eastern edge where vegetation growth between May and August contributes to positive elevation change.

The classified map (b) discretizes these patterns into the 7-class formulation. The High Thaw class (C7, dark red) forms coherent spatial clusters corresponding to active thermokarst features, while the High Heave class (C1, dark blue) dominates the vegetated eastern strip. The strong spatial coherence (Moran's I = 0.99) is visually evident, with similar classes clustering together rather than appearing randomly distributed.

\subsection{Class Imbalance Analysis}

\begin{figure}[h]
\centering
\includegraphics[width=\textwidth]{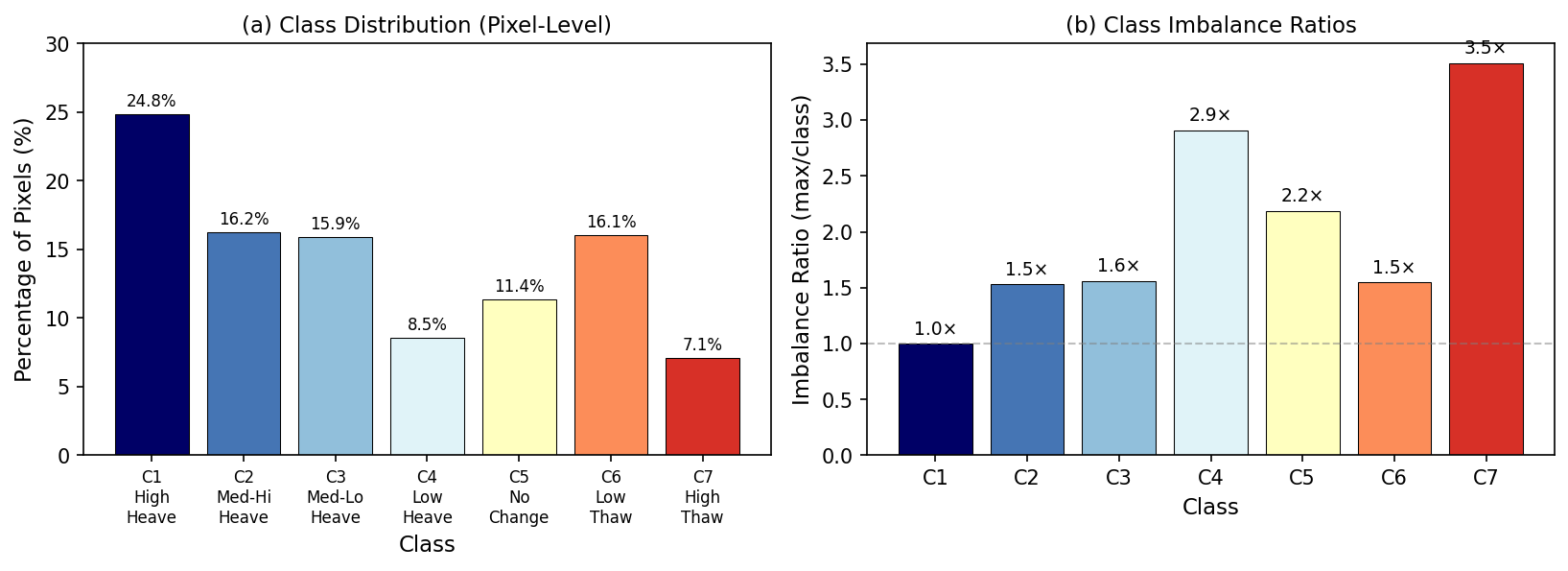}
\caption{Class distribution analysis. (a) Percentage of pixels in each class, showing the heave-dominated nature of the dataset. (b) Imbalance ratios computed as (max class count) / (class count), with the dashed line indicating balanced classes. High Thaw (C7) is 3.5$\times$ underrepresented relative to High Heave (C1).}
\label{fig:imbalance}
\end{figure}

Figure~\ref{fig:imbalance} quantifies the class distribution and resulting imbalance. The bar chart (a) shows that the dataset is heave-dominated, with 65.5\% of pixels falling in heave classes (C1--C4). High Heave (C1) is the most prevalent class at 24.8\%, while High Thaw (C7) is the least common at 7.1\%.

The imbalance ratios (b) reveal moderate class imbalance with a maximum ratio of 3.5$\times$ between the largest (C1) and smallest (C7) classes. This imbalance reflects the physical reality that severe thermokarst subsidence is spatially localized, making accurate detection of High Thaw the most challenging and arguably most important task for infrastructure monitoring. The inverse-frequency weights shown in Table~\ref{tab:class_stats} ensure that the rare High Thaw class (C7) receives proportionally higher loss contributions during training.

\section{Qualitative Results}
\begin{figure}[h]
\centering
\includegraphics[width=\textwidth]{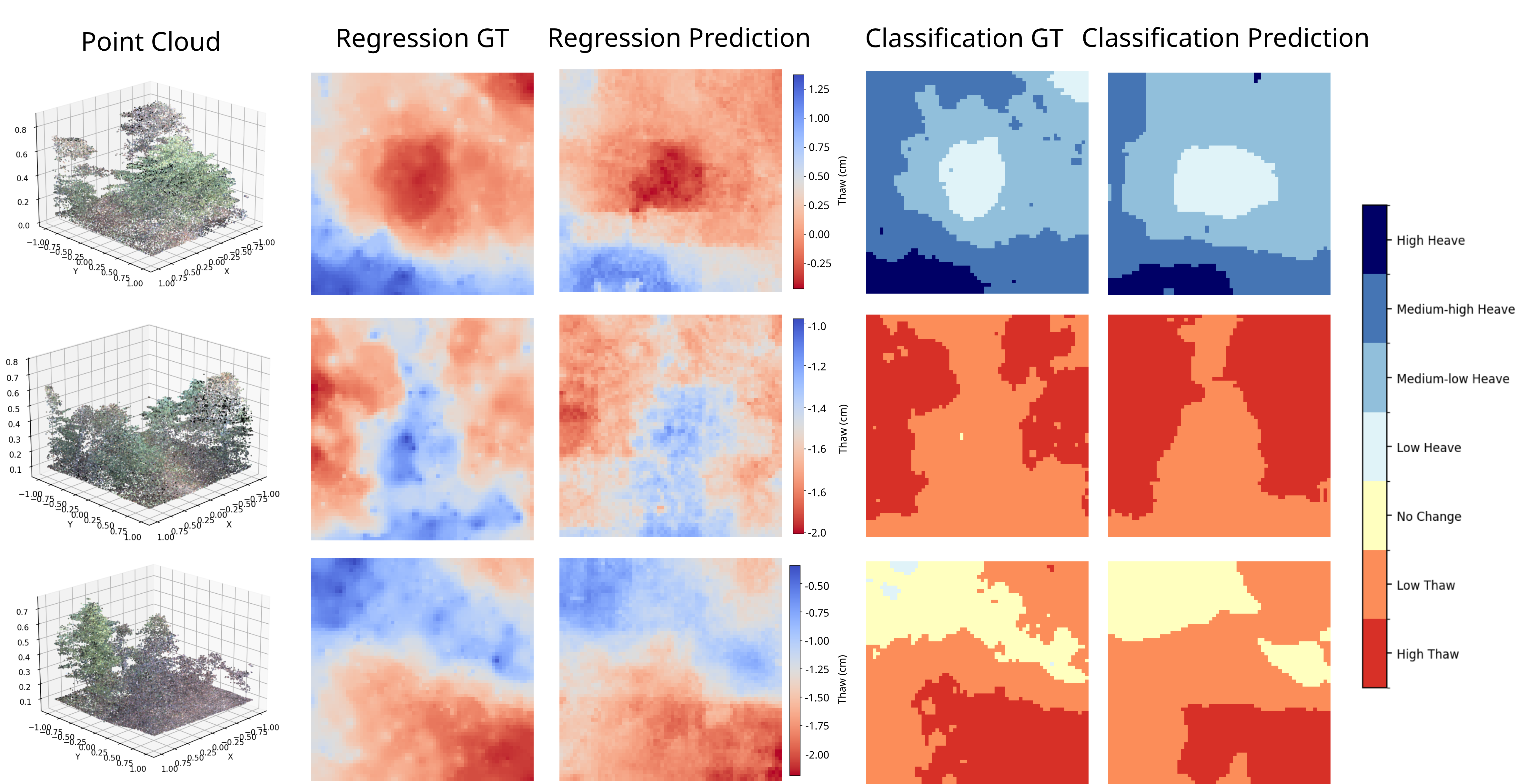}
\caption{Qualitative results on three evaluation tiles. Each row shows (left to right): input point cloud with RGB coloring, regression ground truth, regression prediction, classification ground truth, and classification prediction. Regression colormaps show thaw (red/negative) to heave (blue/positive) in centimeters. Classification legend indicates severity from High Thaw (C7, red) through No Change (C5, cream) to High Heave (C1, dark blue). The model accurately captures spatial patterns of thaw and heave across diverse tile conditions.}
\label{fig:qualitative_2}
\end{figure}

Figure~\ref{fig:qualitative_2} presents qualitative results on three representative evaluation tiles, showing both regression and classification outputs alongside ground truth.

Across all examples, the model demonstrates robust performance in capturing the dominant spatial patterns of permafrost dynamics, with classification predictions that closely match ground truth class boundaries. The regression outputs provide additional granularity for continuous monitoring applications.

